\documentclass{article}

\usepackage{arxiv}

\usepackage[utf8]{inputenc} 
\usepackage[T1]{fontenc}    
\usepackage{hyperref}       
\usepackage{url}            
\usepackage{booktabs}       
\usepackage{amsfonts}       
\usepackage{nicefrac}       
\usepackage{microtype}      
\usepackage{lipsum}		
\usepackage{graphicx}
\usepackage{natbib}
\usepackage{doi}
\usepackage[colorinlistoftodos,prependcaption,textsize=tiny]{todonotes}
\usepackage{threeparttable}
\usepackage[toc,page]{appendix}
\usepackage{multirow}
\usepackage{amsmath}
\usepackage{upgreek}

\newcommand{\dong}[1]{{\color{black}{#1}}} 
\title{HelixVS: Deep Learning–Enhanced Structure-Based Platform for Screening and Design}

\author{ Shanzhuo Zhang, Xianbin Ye, Donglong He, Yueyang Huang, Xiaonan Zhang, Xiaomin Fang\thanks{Corresponding author. Email: fangxiaomin01@baidu.com } \\PaddleHelix Team, Baidu Inc.}

\renewcommand{\headeright}{Technical Report}
\renewcommand{\undertitle}{Technical Report}
\renewcommand{\shorttitle}{HelixVS}

\hypersetup{
pdftitle={A template for the arxiv style},
pdfsubject={q-bio.NC, q-bio.QM},
pdfauthor={David S.~Hippocampus, Elias D.~Striatum},
pdfkeywords={First keyword, Second keyword, More},
}

\begin{document}
\maketitle

\begin{abstract}
Drug discovery through virtual screening (VS) has become a popular strategy for identifying hits against protein targets. \dong{Alongside VS, molecular design further expands accessible chemical space. Together, these approaches have} the potential to reduce the cost and time needed for manual selection and wet-laboratory experiments, thereby accelerating drug discovery pipelines. Improving the cost-effectiveness of virtual screening is a significant challenge, aiming to explore larger compound libraries while maintaining lower screening costs. Here, we present HelixVS, a structure-based VS platform enhanced by deep learning models. HelixVS integrates a precise deep learning-based pose-scoring model and a pose-screening module into a multi-stage VS process, enabling more effective screening of active compounds. Compared to classic molecular docking tools like Vina, HelixVS demonstrated significantly improved screening performance across nearly a hundred targets, achieving an average 2.6-fold higher enrichment factor (EF) and more than 10 times faster screening speed. We applied HelixVS in four drug development pipelines, targeting both traditional competitive drug-binding pockets and novel protein-protein interaction interfaces. Wet-lab validations across these pipelines consistently identified active compounds, with over 10\% of the molecules tested in wet labs demonstrating activity at $\upmu$M or even nM levels. This demonstrates the ability of HelixVS to identify high-affinity ligands for various targets and pockets.\dong{  In addition, the HelixVS platform has been extended with HelixVS-Syn, which enables design of novel compounds from reference scaffolds. These designed molecules are seamlessly integrated into the HelixVS screening workflow, allowing researchers to explore both existing chemical libraries and novel chemical space with high affinity, synthetic accessibility, and structural novelty.} Furthermore, we provide a publicly available and free version of HelixVS with limited computing power to assist drug development scientists in accelerating their drug discovery processes. The HelixVS online service is available at: \url{https://paddlehelix.baidu.com/app/drug/helixvs/forecast}. We also support private deployment solutions to meet the data security requirements of pharmaceutical companies, research institutes, and other organizations. For collaboration inquiries, please contact: \href{mailto:baidubio_cooperate@baidu.com}{baidubio\_cooperate@baidu.com}.


\end{abstract}


\section{Introduction}
Drug discovery is a challenging and complex process that requires extensive knowledge. High-throughput screening (HTS) is a laboratory technique used in drug discovery to identify compounds that show activity against the target of interest. Compared with HTS, a computational approach, virtual screening (VS) vastly accelerates early-stage drug discovery and development due to its cost-effectiveness and time efficiency. Virtual screening is able to explore a large chemical space containing millions to billions of compounds. 

A virtual screening pipeline typically involves three main steps: (1) preparation of the target protein and compound library, (2) molecular docking and scoring to rank candidate compounds, and (3) compound filtering based on expert-defined rules. Among these, step (2) is the key step of the drug discovery pipeline. Conventional molecular docking tools, such as AutoDock, AutoDock Vina \cite{trott2010autodock,doi:10.1021/acs.jcim.1c00203}, and Glide \cite{friesner2004glide,halgren2004glide}, are used to predict ligand binding modes and affinities to target proteins. These tools are founded on the physical principles of molecular interactions, using empirical and/or physics-based scoring functions to evaluate the binding affinities, and can effectively narrow down the number of potential drug candidates and prioritize those with the highest probability of success. However, they are limited by the accuracy of their scoring functions, which can result in wrong docking poses and faulty active or decoy decisions.

On the other hand, advanced deep learning-based drug-target affinity (DTA) prediction models, especially structure-based models \cite{li2021structure} \cite{jimenez2018k} \cite{jiang2021interactiongraphnet} \cite{moon2022pignet} \cite{rtmscore2022}, can provide more accurate affinity predictions. They learn from large datasets of known target-ligand interactions to provide accurate predictions of binding affinities. However, the robustness of DTA model performance across different protein targets cannot be guaranteed, as the trained models may suffer from overfitting, where the model performs well on training data but fails to generalize to new data. Since both docking tools and DTA models have their strengths and weaknesses, it is attractive to integrate these two technologies into different stages of the virtual screening pipeline to take advantage of their respective benefits and improve screening accuracy.

Based on this understanding, we developed HelixVS, a cost-effective virtual screening platform to enhance screening efficiency and success rate in early-stage drug discovery. HelixVS performs multi-stage screening, comprehensively taking advantage of both molecular docking tools and deep learning-based models to screen out potential active molecules against target proteins. \dong{Moreover, the platform is extended with HelixVS-Syn, a module that enables scaffold-based molecular design, further broadening the accessible chemical space before screening.} HelixVS demonstrated significantly improved screening performance compared to using only molecular docking tools. Compared to Vina across over a hundred targets in the DUD-E dataset \cite{dude2012jmc}, HelixVS can find an average of 159\% more active molecules and run up to nearly 15 times faster. These results highlight the effectiveness of HelixVS's multi-stage screening process.

Our application of the HelixVS platform in four real drug development pipelines exemplifies this approach. The targets, ranging from cyclin-dependent kinases CDK4/6 and NF-$\upkappa$B inducing kinase (NIK) to immune modulators TLR4/MD-2 and cGAS, represent diverse mechanisms of disease pathogenesis and potential therapeutic intervention. The use of HelixVS enabled the screening of libraries containing millions of small molecules, identifying not only potential inhibitors with significant activity at $\upmu$M or even nM concentrations but also facilitating the exploration of novel protein-protein interaction interfaces.

The HelixVS platform offers a virtual screening service for drug development scientists, providing a comprehensive virtual screening service whose key features can be summarized as follows:
\begin{itemize}
    \item \textbf{Superior Hit Rate:} HelixVS significantly improves virtual screening hit rates through multi-stage screening strategies and integration of deep learning models. Testing on the DUD-E dataset shows that HelixVS can achieve enrichment factors (EF) of 44.205 and 26.968 at 0.1\% and 1\% respectively, significantly outperforming other methods. In multiple actual drug development projects, including difficult scenarios such as dual-target and protein-protein interface (PPI) binding, HelixVS has successfully identified active molecules.\dong{HelixVS-Syn further complements this capability. Under the same computational budget, HelixVS-Syn designs molecules that surpass those from the VS in synthetic accessibility, binding affinity, and novelty.} These results demonstrate HelixVS platform's exceptional capability in identifying active compounds.
	\item \textbf{Excellent Screening Throughput and Cost-effectiveness:} HelixVS leverages Baidu Cloud's CPU computing infrastructure\footnote{https://cloud.baidu.com} and CHPC's high-performance computing resources\footnote{https://cloud.baidu.com/product/chpc.html} to achieve screening throughput exceeding 10 million molecules per day. The platform's resource dynamic scaling capabilities and multi-stage screening workflow enable cost-effective screening across different computing resource scales, with screening costs reduced to as low as 1 RMB per thousand molecules.
    \item \textbf{Enhanced Usability:} HelixVS offers automated protein pre-processing, large-scale built-in compound library, custom compound library, and binding mode filtering abilities. These features are accessible through a straightforward web interface, where users can input parameters easily. This functionality is particularly beneficial for medicinal chemists who are not familiar with computational tools, allowing them to complete real virtual screening tasks quickly and accurately. The screenshots of the HelixVS web interface are shown in Appendix Section \ref{sec:screenshots}.
\end{itemize}

\section{The HelixVS Platform}

\begin{figure}
\centering
\includegraphics[width=1.0\columnwidth]{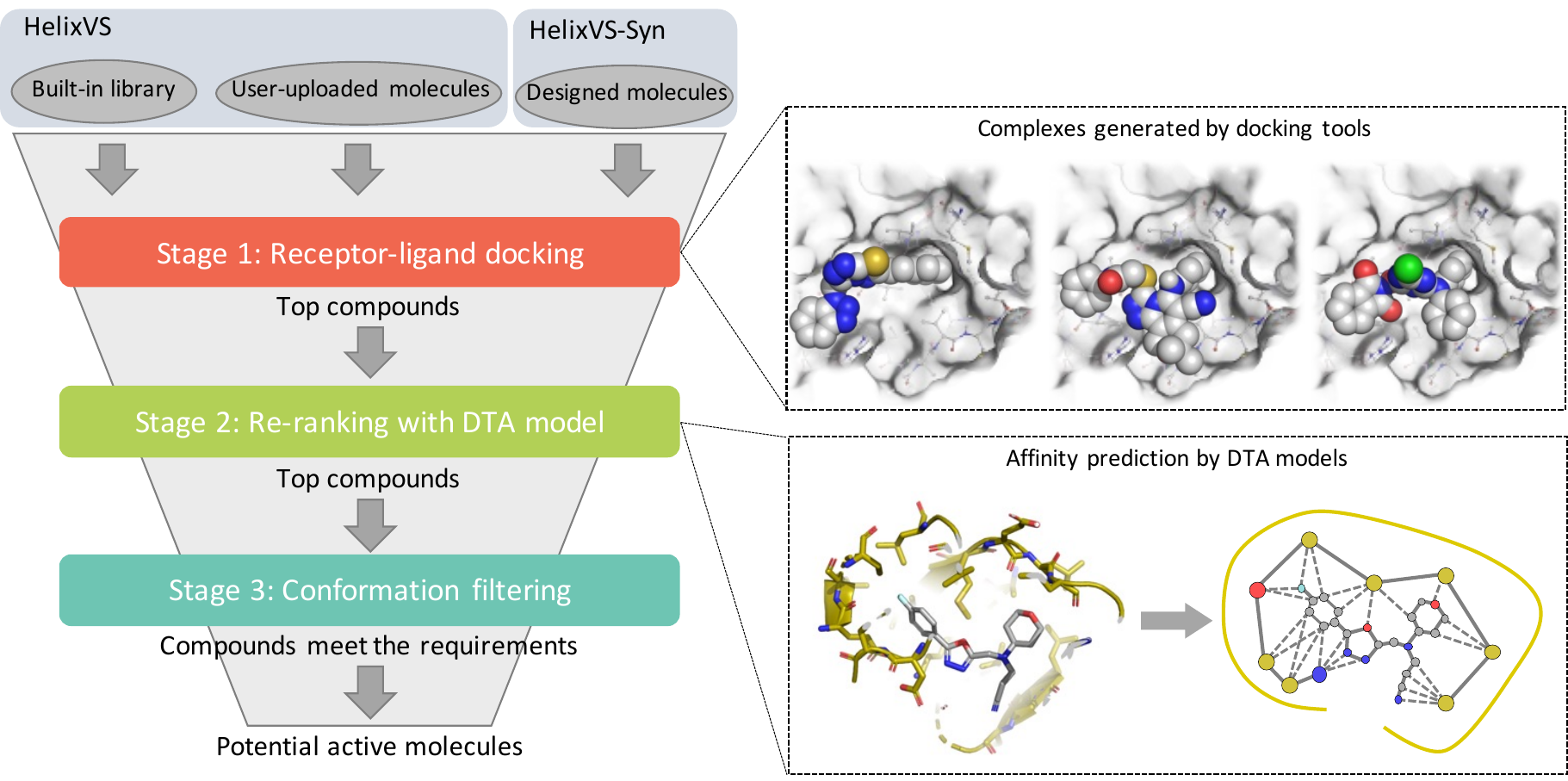}
\caption{Overall pipeline of the HelixVS platform.}
\label{fig:helixvs_overall_pipeline}
\end{figure}
\dong{The HelixVS platform supports multiple sources of molecular inputs, including its built-in large-scale compound library, user-defined molecular libraries, and newly designed molecules generated by HelixVS-Syn from reference scaffolds. All these inputs can be seamlessly integrated into the subsequent three-stage drug discovery pipeline, ensuring that diverse molecular sources undergo the same systematic screening to identify potential active compounds against specific protein targets.(as shown in Figure~\ref{fig:helixvs_overall_pipeline}).} 
\subsection{Multi-staged virtual screens with HelixVS}
In Stage 1, HelixVS employs classical docking tools that rely on scoring functions and sampling techniques to generate drug-target binding poses. By default, we utilize AutoDock QuickVina 2 for docking \cite{alhossary2015fast}, as it offers significantly faster speed compared to AutoDock Vina \cite{trott2010autodock}. Although its scoring function is simpler and may retain non-optimal molecular conformations, we address this limitation by preserving multiple binding conformations after docking. Through our testing, we have found that retaining multiple conformations for the next stage can substantially improve enrichment factors on test datasets. Following the docking stage, molecules with higher affinity scores are selectively retained and enter the next stage.

In Stage 2, the docking poses with lower $\Delta G$ are fed into a deep learning-based affinity scoring model to obtain more accurate binding conformation scores compared to docking tools.
In HelixVS, the main structure of our model is based on RTMscore, which has been proven to have high accuracy in small molecule screening scenarios \cite{rtmscore2022}. Building upon this foundation, we collected additional co-crystal structure data, covering almost all ligands and their spatial structures in the PDB database, to enhance RTMscore through data augmentation. At this stage, HelixVS simultaneously considers multiple isomers and uses multiple docking conformations, thereby increasing the likelihood of finding optimal conformations and affinity scores.

In Stage 3, HelixVS introduces an optional conformation filtering step to further refine molecules with higher scores based on pre-defined binding modes. This step enables targeted screening for molecules that specifically bind to certain amino acids with specific interactions. Subsequently, the remaining molecules are clustered, and representative molecules are selected to ensure diversity of the screening results.

Between each stage, HelixVS employs highly efficient distributed sorting algorithms to rank the current molecular list and gradually filter out superior molecules.

\subsection{Virtual Screening Performance on DUD-E}

To demonstrate the performance of HelixVS, we tested the screening power (i.e. the ability to find active molecules among a pool of active and decoy molecules) on the DUD-E \cite{dude2012jmc} dataset. DUD-E is one of the most popular virtual screening benchmarks, containing 102 proteins from 8 diverse protein families and a rich repository of 22,886 active molecules. It curates 50 topologically distinct decoys for each active molecule that possess matched physicochemical properties, sourced from the extensive ZINC database.

Here, we compare the performance of our HelixVS with other methods on DUD-E, including classic docking software Vina \cite{trott2010autodock} and Glide SP \cite{friesner2004glide}, as well as the recently proposed deep learning-based docking model KarmaDock \cite{KarmaDock2023NCS}. Implementation details are reported in Section \ref{exp_setting}. All methods mentioned above were evaluated on two commonly used metrics, enrichment factor \cite{EFLi2009} and screening speed. The overall performance is depicted in Table \ref{tab:dude_overall_performance}. We can see from the table that HelixVS outperforms commonly used classic docking softwares (eg. Vina and Glide SP) by a large margin on both enrichment factor and running time. Compared to Vina, HelixVS can find an average of 159\% more active molecules and run up to nearly 15 times faster. Besides, HelixVS shows an approximate 70.3\% and 16.8\% improvement in EF at 0.1\% respectively, compared with the deep learning-based molecular docking model Karmadock and commercial software Glide SP.

\begin{table}[ht]
\centering
\begin{threeparttable}
    \caption{Overall performance and speed of HelixVS and other tools on DUD-E.}
    \centering
    \begin{tabular}{lcccc}
        \toprule
              \multirow{2}{*}{Method} & \multirow{2}{*}{$\mathrm{EF}_{0.1\%}$} & \multirow{2}{*}{$\mathrm{EF}_{1\%}$}  & \multicolumn{2}{c}{Screening Speed (molecules / day)} \\
                \cmidrule{4-5}
                & & & per CPU core & per GPU card\\
        \midrule
            Vina  & 17.065 & 10.022 & $\sim$300 $^{a}$  & $-$ \\ 
            Glide SP & 37.842 & 24.346 & $\sim$2400 $^{b}$  & $-$ \\
            KarmaDock & 25.958 & 15.848 & $-$ & $\sim$5 Million $^{c}$ \\
            HelixVS & \textbf{44.205} & \textbf{26.968} & $\sim$4000 & $-$ \\
        \bottomrule
    \end{tabular}
    \label{tab:dude_overall_performance}
    \begin{tablenotes}
        \footnotesize
        \item[a] The screening speed is adapted from \cite{VirtualFlow2020}, with \textit{exhaustiveness} = 32.
        \item[b] The screening speed is reported by \cite{GlideSP_speed}.
        \item[c] The screening speed is tested on NVIDIA V100 32G GPU. As reported by \cite{KarmaDock2023NCS}.
    \end{tablenotes}
\end{threeparttable}
\end{table}

We also investigated the performance of all methods on each individual target families, detailed results as reported in Appendix \ref{Appendix_dude}. Surprisingly, HelixVS performs even better for certain protein categories, such as kinases, proteases, nuclear receptors, ion channels, etc. Concretely, on an important target of Parkinson's disease, COMT (Catechol-O-methyltransferase) \cite{COMT2014}, HelixVS is the only algorithm that can find active molecules, with an EF~~0.1 as high as 61.51. 

Recently, a new deep learning-based molecular docking model named Boltz-2 \cite{passaro_boltz-2_2025} has been proposed. It is a binary classification model that predicts whether a molecule can bind to a protein target based on its 3D structure. We also tested the performance of Boltz-2 on DUD-E. However, due to the significantly higher GPU resource requirements of Boltz inference compared to other methods, we selected 10 targets with smaller total ligand counts in DUD-E for inference and evaluation (specific selected targets and molecule counts are detailed in Appendix Table \ref{tab:dude_10_targets}). The results are reported in Table \ref{tab:dude_boltz}. Boltz-2 has two different scoring modes: in binary mode, its EF 0.1 is lower than HelixVS, while EF 1 is higher than HelixVS; in affinity mode, both EF 0.1 and EF 1 are significantly lower than HelixVS. When considering computational costs, given that the cost of Nvidia GPUs with over 40GB VRAM can be 100-200 times that of a single CPU core, HelixVS's total cost is estimated to be only 1/500 of Boltz-2's, demonstrating HelixVS's high cost-effectiveness in virtual screening scenarios.

\begin{table}[ht]
	\centering
	\begin{threeparttable}
		\caption{Overall performance and speed of HelixVS, Boltz-2 and other tools on 10 targets of DUD-E.}
		\centering
		\begin{tabular}{lcccc}
			\toprule
				  \multirow{2}{*}{Method} & \multirow{2}{*}{$\mathrm{EF}_{0.1\%}$} & \multirow{2}{*}{$\mathrm{EF}_{1\%}$}  & \multicolumn{2}{c}{Screening Speed (molecules / day)} \\
					\cmidrule{4-5}
					& & & per CPU core & per GPU card\\
			\midrule
				Vina  & 10.948 & 9.308 & $\sim$300 & $-$ \\ 
				Glide SP & 24.492 & 21.387 & $\sim$2400 & $-$ \\
				KarmaDock & 8.924 & 4.104 & $-$ & $\sim$5 Million \\
				Boltz-2 binary  & 33.767 & \textbf{27.753} & $-$ & $\sim$1440 $^{a}$ \\ 
				Boltz-2 affinity & 19.500 & 14.281 & $-$ & $\sim$1440 $^{a}$ \\
				HelixVS & \textbf{35.306} & 24.413 & $\sim$4000 & $-$ \\
			\bottomrule
		\end{tabular}
		\label{tab:dude_boltz}
		\begin{tablenotes}
			\footnotesize
			\item[a] The screening speed is tested on NVIDIA A800 80G GPU. This test measures the inference speed excluding MSA retrieval, as MSA results can be reused in screening scenarios. The Boltz-2 speed is calculated based on approximately 1 minute required for inference per molecule.
		  \end{tablenotes}
		  \end{threeparttable}
	\end{table}

To summarize, all of the above results demonstrate that HelixVS is indeed an effective virtual screening tool compared to classical docking software and deep learning-based docking model.

\subsection{Compound Library Pre-processing}

HelixVS supports virtual screening with both built-in compound libraries and user-uploaded custom molecular libraries, providing flexibility for diverse drug discovery needs. The built-in libraries consist of four different scales of small molecule libraries, ranging from tens of thousands to tens of millions of compounds, sourced from TopScience\footnote{\url{https://www.targetmol.cn/topscience-database}}, a leading provider of chemical compounds. This ensures that the majority of molecules in our built-in libraries are commercially available and purchasable for follow-up experimental validation. The screening time and price of built-in libraries are reported in Appendix Table \ref{tab:compound_libraries}. For custom molecular libraries, users can upload their own compound collections in SMILES format, enabling personalized virtual screening campaigns tailored to specific research requirements. To ensure structural rationality and computational efficiency, all compounds (both built-in and custom) undergo rigorous preprocessing steps including: (1) desalting and neutralization to standardize molecular forms, (2) stereoisomer and tautomer enumeration to account for all possible molecular conformations, (3) protonation to reflect physiological conditions, and (4) structure generation and energy filtering to eliminate energetically unfavorable conformations. This comprehensive preprocessing pipeline ensures that only high-quality, drug-like molecules with reasonable structural properties are included in the screening process, thereby improving the overall success rate of virtual screening campaigns.

\subsection{Design Novel Molecules with HelixVS-Syn}
\dong{HelixVS-Syn enables design of novel compounds from a known molecular scaffold and integrates these designs into a complete virtual-screening workflow. It is a pathway-driven molecular design algorithm that starts from synthesizable and purchasable compounds. By integrating retrosynthetic knowledge with synthesis-guided modifications, it generates new molecules that remain synthetically accessible. This enables exploration of a broad chemical space while reducing computational cost and research time compared with large-scale screening.

We conducted a comparative experiment against HelixVS performed on a compound library of 820,000 molecules, using CDK5 as the target protein. From the 50,000 molecules generated by HelixVS-Syn, we selected the top 50 for statistical analysis. 

Compared to HelixVS, HelixVS-Syn produces compounds with higher synthetic accessibility, affinity and novelty(how these metrics are calculated is reported in Appendix Section\ref{Appendix_syn}).}
\begin{table}[h]
\centering
\begin{tabular}{lccc}
\hline
Method & Synthetic accessibility & Affinity & Novelty \\
\hline
HelixVS-Syn      & \textbf{0.641} & \textbf{60.69} & \textbf{0.708} \\
HelixVS        & 0.615 & 64.54 & 0.658 \\
\hline
\end{tabular}
\caption{HelixVS-Syn performance on CDK5 compared with HelixVS}
\label{tab:syn_methods_comparison}

\end{table}

\section{HelixVS in Real Drug Discovering Pipelines}

\subsection{Screening of CDK4/6 Dual-target Inhibitors with HelixVS}

Cyclin-dependent kinases 4 and 6 (CDK4/6) are key regulators of the cell cycle, controlling the progression of cells from the G1 phase to the S phase where DNA replication occurs. CDK4/6 activity is tightly regulated by the binding of cyclin D1 (CCND1), which activate the kinase domain and initiate downstream signaling pathways \cite{oleary_treating_2016}. Dysregulation of CDK4/6 has been implicated in the development and progression of many human cancers, making it an important target for cancer therapy. However, existing CDK4/6 inhibitors have limited specificity, leading to off-target effects and toxicity \cite{liu_breaking_2025}. In addition, cancer cells can develop resistance to these inhibitors through mutations in the ATP binding pocket or upregulation of alternative signaling pathways.

Targeting the CDK4/6-CCND1 interaction represents an alternative and more favorable strategy for inhibiting CDK4/6 activity. By preventing the formation of active CDK4/6-CCND1 complexes, these inhibitors can selectively target cancer cells with high levels of CCND1 expression, while sparing normal cells with low CCND1 expression. In addition, these inhibitors may be less susceptible to resistance mechanisms involving mutations in the ATP binding pocket \cite{morin_oncogene_2000}. 

Aiming at discovering a lead compound targeting CDK4/6-CCND1 interaction, HelixVS platform was used to screen a 7.8 million compound library for potential molecules that could bind to a previously unreported pocket on the CDK4/6-CCND1 interface. From the final set of 100 compounds, we purchased 40 commercially available compounds for further activity evaluation. We used an unpublished method named the bimolecular fluorescence complementation (BiFC) assay to confirm the activity of these compounds against CDK4/6-CCND1 interaction (details in the Methods section). The BiFC assay identified 6 compounds disrupting the formation of CDK4/6-CCND1 complex, as evidenced by a more than 20\% decrease in fluorescence intensity. Our results demonstrate the accuracy and powerfulness of HelixVS in hit identification, underscores the platform's ability to tackle complicated cases such as the discovery of PPI inhibitors. We believe HelixVS has a great potential to make significant contribution to the development of novel therapeutics and ultimately benefit patients in need.


\subsection{Screening of TLR4/MD-2 Inhibitors with HelixVS}

Toll-like receptor 4 (TLR4) from the Toll-like receptor family is an important pattern recognition receptor that plays a key role in the immune system, especially in innate immune responses. 
It is physically associated with MD-2 and can recognize the major component of bacterial cell walls, LPS, activate downstream signaling pathways, lead to the production of inflammatory factors, and initiate inflammatory responses \cite{shimazuMD2MoleculeThat1999}. Research suggests that inhibiting the activation of TLR4 may help reduce inflammatory responses in the treatment of autoimmune diseases such as rheumatoid arthritis and systemic lupus erythematosus \cite{liu_tlr2_2014}. In addition, the overactivation of TLR4 also appears in some infectious diseases, allergic diseases, and cancers, making TLR4 a potential target for treatment strategies of many diseases \cite{giat_cancer_2017}.

However, as of now, research on TLR4 as a drug target is mainly focused on the preclinical and clinical trial phases, and many of them are macromolecular drugs. We are committed to developing new small molecule inhibitors targeting the innovative pocket on the PPI surface between TLR4 and MD-2, which are more convenient for oral administration. Utilizing HelixVS, we meticulously screened a library of 200k molecules and selected 103 candidates for wet-lab experiments. These molecules were assessed for activity using the SEAP assay \cite{TLR42009}, which led to the identification of 6 highly potential candidates, 2 of which demonstrated nM-level activity. Follow-up modification and molecular dynamics validation of one of these molecules have been published \cite{jiang_discovery_2024}.


\subsection{Screening of cGAS Inhibitors with HelixVS}

The cyclic GMP-AMP synthase (cGAS) acts as a cytoplasmic enzyme activated by viral infections or DNA damage, triggering downstream immune responses. Consequently, inhibiting cGAS or the STING pathway is being explored as a therapeutic strategy for autoimmune diseases like systemic lupus erythematosus and rheumatoid arthritis \cite{hu_emerging_2022}. In this project, targeting the ATP-binding pocket of cGAS, we utilized HelixVS to screen a smaller library of 30,000 compounds and identified 126 candidates. A cell-based lucia luciferace assay \cite{lamaDevelopmentHumanCGASspecific2019} validated 96 molecules, revealing 17 with significant activity. Remarkably, 10 of these molecules demonstrated activity below 10 $\upmu$M, and one demonstrated nM-level activity.



\subsection{Screening of NIK Inhibitors with HelixVS}

NF-$\upkappa$B inducing kinase (NIK), also known as MAP3K14, is a crucial regulatory kinase in the non-canonical NF-$\upkappa$B signaling pathway. Clinical observations have linked the overactivation of NIK to the pathogenesis of inflammatory diseases, B-cell malignancies, and solid tumors, positioning NIK inhibition as an appealing drug discovery strategy. Such inhibitors could potentially treat conditions like cancer, inflammatory disorders, metabolic dysfunctions, and autoimmune diseases \cite{pflugTargetingNFkBInducingKinase2020}. In this project, targeting the ATP-binding pocket of NIK, we utilized HelixVS to conduct a virtual screening of approximately 10 million compounds from a consolidated library to discover new active molecular scaffolds. The top-ranked molecules underwent molecular structure-based clustering, from which 7 candidates were selected for enzymatic activity assays, revealing one molecule with an IC50 in the $\upmu$M range.




\section{Conclusion}
The integration of HelixVS into our drug discovery pipeline represents a significant advancement in the quest for new therapeutic agents. By utilizing this sophisticated virtual screening platform, we have successfully identified potential inhibitors that target a variety of critical proteins involved in disease mechanisms, such as CDK4/6-CCND1, TLR4/MD-2, cGAS, and NIK. These findings highlight the platform's robust capability to navigate complex molecular landscapes and pinpoint candidates with promising biological activity, some exhibiting potent inhibitory effects at $\upmu$M or even nM concentrations.

Our studies have demonstrated that HelixVS not only enhances the efficiency of the screening process but also significantly improves the success rate by leveraging both classical molecular docking tools and advanced deep learning-based models. This dual approach ensures a comprehensive exploration of chemical space and a higher likelihood of discovering viable hit compounds. Furthermore, the platform's ability to handle vast libraries and deliver rapid screening results has been instrumental in accelerating the early stages of drug development.\dong{Notably, the integration of HelixVS-Syn further extends this capability by enabling scaffold-based molecular design, thereby expanding the accessible chemical space with novel, synthetic accessibility, and active candidates.}


\clearpage

\begin{appendices}
\renewcommand{\thesection}{\arabic{section}}

\counterwithin{figure}{section}
\counterwithin{table}{section}
\counterwithin{equation}{section}

\section{Detailed Experimental Settings}
\subsection{Screening Powers Evaluation}
\label{exp_setting}
    The docking protocol for Vina is as follows: the docking box is centered at the geometric center of the native ligand with a 15 Å length cubic region, followed by the standard docking pipeline, executed on CPU (Intel(R) Xeon(R) Gold 6248 CPU @ 2.50GHz) with its default scoring function and parameters except that the \textit{exhaustiveness} is set to 32. For Glide SP, all results including docking poses and scores are freely accessibl docking poses and scores are freely accessible through \cite{rtmscore2022}. For Karmadock, wFor KarmaDock, we load the trained model parameters provided by KaramaDock \cite{KarmaDock2023NCS} and follow their settings to report the results, using NVIDIA V100 32G GPU.

\subsection{The Bimolecular Fluorescence Complementation Assay}

The BiFc assay was used to detect protein-protein interactions between CDK4/6 and CCND1. Briefly, the sequences of VN155-CDK4/6 and VC155-CCND1 were cloned into the vectors carrying the N- and C-terminal halves of the green fluorescent protein (GFP), respectively. The constructed vectors were confirme by sequencing. HEK293T cells were grown to 50\% confluency and co-transfected with VN155-CDK4/VC155-CCND1 or VN155-CDK6/VC155-CCND1 vectors. The PPI fluorescence signal was detected by fluorescence microscopy after 24 hours (Figure). To evaluate the inhibitory activities of the tested compounds, transfected cells were treated with compounds at a final concentration of 10 \textmu M for 18 hours. The fluorescence intensity was quantified using a high-content imaging system. Cells treated with DMSO was used as a blank. The mCherry signal intensity was used as a reference signal for normalizing the PPI signal.

\subsection{Molecule Synthetic accessibility, Affinity and Novelty Evaluation Metric}
\label{Appendix_syn}
\dong{
Synthetic accessibility was evaluated by the Retro*\cite{chen2020retrolearningretrosyntheticplanning} success rate. The Retro* success rate is defined as the proportion of molecules in a set for which Retro* can identify a synthetic route. Affinity was assessed using calculated affinity scores.
\[
\text{Synthetic accessibility} = \frac{1}{N} \sum_{i=1}^{N} I_i
\]

\noindent where
\begin{tabular}{ll}
$N$ &: total number of molecules under evaluation, \\
$I_i$ &: indicator function: $I_i = 1$ if Retro* identifies a valid synthetic route for molecule $i$, and $I_i = 0$ otherwise. \\
\end{tabular}

Affinity was predicted by the DTA models.

Novelty was assessed by computing the Tanimoto similarity between the generated molecules and a reference compound library of 200,000 molecules. For each generated molecule, the mean similarity score to its 50 most similar molecules in the library was calculated. The overall novelty score was then obtained as the mean of these similarity values across all generated molecules, and defined as \[
\text{Novelty} = 1 - \frac{1}{N} \sum_{i=1}^{N} 
\left( \frac{1}{50} \sum_{j=1}^{50} S_{i,j} \right)
\]

\noindent where
\begin{tabular}{ll}
$N$ &: total number of generated molecules, \\
$S_{i,j}$ &: Tanimoto similarity between the $i$-th molecule and its $j$-th most similar compound in the reference library\\
$\tfrac{1}{50}\sum_{j=1}^{50} S_{i,j} $ &: mean similarity of the $i$-th molecule to its 50 most similar library molecules, \\
$\tfrac{1}{N}\sum_{i=1}^{N}(\cdot)$ &: mean similarity across all generated molecules. \\
\end{tabular}
}
\section{Detailed Results on DUD-E}

\subsection{Virtual Screening Performance about each individual target family on DUD-E}

We report the performance (EF~0.1\% and EF~1\%) of each method on an individual target family, as shown from table \ref{tab:dude_Cytochrome_P450} to \ref{tab:dude_Protease}. In short, HelixVS achieved state-of-the-art results on EF~~0.1\% on six of the eight target families, compared to the other methods. Especially in the Ion Channel target family, HelixVS surpasses all other baselines by a large margin, achieving a 5.847-fold and 2.560-fold improvement compared with Vina and Glide SP, respectively. Unfortunately, due to the complexity of the binding mechanism of the GPCR and Cytochrome P450 targets, HelixVS performed slightly less effectively than the best methods, finishing as the runner-up. Considering its robust performance, we can expect the promising prospect of HelixVS in real-world virtual screening tasks.

\label{Appendix_dude}
    \begin{table}[ht]
    \begin{minipage}[b]{0.3\textwidth}
        \begin{threeparttable}
    	\caption{Cytochrome P450.}
    	\centering
    	\begin{tabular}{lcc}
    		\toprule
    		      Method & $\mathrm{EF}_{0.1\%}$ & $\mathrm{EF}_{1\%}$ \\
    		\midrule
        		Vina  & 3.256 & 3.607  \\
                    Glide SP & 6.399 & 3.923 \\
                    KarmaDock & 12.732 & 4.347 \\
                    HelixVS & 6.512 & 4.811 \\
    		\bottomrule
    	\end{tabular}
    	\label{tab:dude_Cytochrome_P450}
          \end{threeparttable}
    \end{minipage}%
    \hspace{3mm}%
    \begin{minipage}[b]{0.3\textwidth}
        \begin{threeparttable}
    	\caption{GPCR.}
    	\centering
    	\begin{tabular}{lcc}
    		\toprule
    		      Method & $\mathrm{EF}_{0.1\%}$ & $\mathrm{EF}_{1\%}$ \\
    		\midrule
        		Vina  & 3.966 & 2.186  \\
                    Glide SP & 31.555 & 19.129 \\
                    KarmaDock & 6.298 & 1.736 \\
                    HelixVS & 23.190 & 13.616 \\
    		\bottomrule
    	\end{tabular}
    	\label{tab:dude_GPCR}
          \end{threeparttable}
    \end{minipage}%
    \hspace{3mm}%
    \begin{minipage}[b]{0.3\textwidth}
    \begin{threeparttable}
    	\caption{Ion Channel.}
    	\centering
    	\begin{tabular}{lcc}
    		\toprule
    		      Method & $\mathrm{EF}_{0.1\%}$ & $\mathrm{EF}_{1\%}$ \\
    		\midrule
        		Vina  & 10.348 & 7.734  \\
                    Glide SP & 19.905 & 14.200 \\
                    KarmaDock & 11.007 & 1.122 \\
                    HelixVS & 70.854 & 42.150 \\
    		\bottomrule
    	\end{tabular}
    	\label{tab:dude_Ion_Channel}
          \end{threeparttable}
    \end{minipage}
    \end{table}

    \begin{table}[ht]
    \begin{minipage}[b]{0.3\textwidth}
        \begin{threeparttable}
    	\caption{Kinase.}
    	\centering
    	\begin{tabular}{lcc}
    		\toprule
    		      Method & $\mathrm{EF}_{0.1\%}$ & $\mathrm{EF}_{1\%}$ \\
    		\midrule
        		Vina  & 22.453 & 13.118  \\
                    Glide SP & 48.765 & 28.640 \\
                    KarmaDock & 45.624 & 30.565 \\
                    HelixVS & 51.890 & 30.183 \\
    		\bottomrule
    	\end{tabular}
    	\label{tab:dude_Kinase}
          \end{threeparttable}
    \end{minipage}%
    \hspace{3mm}%
    \begin{minipage}[b]{0.3\textwidth}
        \begin{threeparttable}
    	\caption{Miscellaneous.}
    	\centering
    	\begin{tabular}{lcc}
    		\toprule
    		      Method & $\mathrm{EF}_{0.1\%}$ & $\mathrm{EF}_{1\%}$ \\
    		\midrule
        		Vina  & 4.002 & 10.887  \\
                    Glide SP & 33.325 & 26.342 \\
                    KarmaDock & 8.971 & 3.735 \\
                    HelixVS & 41.325 & 23.661 \\
    		\bottomrule
    	\end{tabular}
    	\label{tab:dude_Miscellaneous}
          \end{threeparttable}
    \end{minipage}%
    \hspace{3mm}%
    \begin{minipage}[b]{0.3\textwidth}
    \begin{threeparttable}
    	\caption{Nuclear Receptor.}
    	\centering
    	\begin{tabular}{lcc}
    		\toprule
    		      Method & $\mathrm{EF}_{0.1\%}$ & $\mathrm{EF}_{1\%}$ \\
    		\midrule
        		Vina  & 27.834 & 13.672  \\
                    Glide SP & 48.071 & 29.355 \\
                    KarmaDock & 16.227 & 8.752 \\
                    HelixVS & 48.518 & 25.617 \\
    		\bottomrule
    	\end{tabular}
    	\label{tab:dude_Nuclear_Receptor}
          \end{threeparttable}
    \end{minipage}
    \end{table}

    \begin{table}[ht]
    \centering%
    \begin{minipage}[b]{0.48\textwidth}
    \centering%
        \begin{threeparttable}
    	\caption{Other Enzymes.}
    	\centering
    	\begin{tabular}{lcc}
    		\toprule
    		      Method & $\mathrm{EF}_{0.1\%}$ & $\mathrm{EF}_{1\%}$ \\
    		\midrule
        		Vina  & 16.374 & 8.933  \\
                    Glide SP & 27.571 & 20.084 \\
                    KarmaDock & 13.459 & 8.056 \\
                    HelixVS & 36.031 & 25.300 \\
    		\bottomrule
    	\end{tabular}
    	\label{tab:dude_Other_Enzymes}
          \end{threeparttable}
    \end{minipage}%
    \hspace{2mm}%
    \begin{minipage}[b]{0.48\textwidth}
    \centering%
        \begin{threeparttable}
    	\caption{Protease.}
    	\centering
    	\begin{tabular}{lcc}
    		\toprule
    		      Method & $\mathrm{EF}_{0.1\%}$ & $\mathrm{EF}_{1\%}$ \\
    		\midrule
        		Vina & 13.307 & 8.081  \\
                    Glide SP & 46.242 & 28.609 \\
                    KarmaDock & 44.980 & 26.479  \\
                    HelixVS & 57.286 & 32.871  \\
    		\bottomrule
    	\end{tabular}
    	\label{tab:dude_Protease}
          \end{threeparttable}
    \end{minipage}
    \end{table}

\begin{table}[ht]
\centering
\caption{10 targets selected from DUD-E for Boltz-2 evaluation}
\begin{tabular}{lccc}
\toprule
Target Name & Class & \#Actives & \#Decoys \\
\midrule
AMPC & Other Enzymes & 48 & 2850 \\
COMT & Other Enzymes & 41 & 3850 \\
GLCM & Other Enzymes & 54 & 3800 \\
INHA & Other Enzymes & 43 & 2300 \\
KITH & Other Enzymes & 57 & 2850 \\
PUR2 & Other Enzymes & 50 & 2700 \\
PYGM & Other Enzymes & 77 & 3950 \\
SAHH & Other Enzymes & 63 & 3450 \\
FABP4 & Miscellaneous & 47 & 2750 \\
CXCR4 & GPCR & 40 & 3406 \\
\bottomrule
\end{tabular}
\label{tab:dude_10_targets}
\end{table}

\section{Built-in compound libraries and screening pricing}

\begin{table}[ht]
\centering
\begin{threeparttable}
\caption{Built-in compound libraries and screening pricing}
\begin{tabular}{lccc}
\toprule
Library Name & \#Molecules & Screening Time (h) $^{a}$ & Screening Price (RMB) \\
\midrule
Targetmol\_CherryPick & $\sim$34,000 & $<$1 & 70 \\
Lifechemicals & $\sim$820,000 & $<$6 & 1,200 \\
ChemDiv & $\sim$2,830,000 & $<$11 & 3,700 \\
Topscience database & $\sim$16,870,000 & $<$26 & 19,000 \\
\bottomrule
\end{tabular}
\label{tab:compound_libraries}
\begin{tablenotes}
    \footnotesize
    \item[a] Screening time when resources are uncontested, may occur when multiple "Topscience database" tasks are running.
\end{tablenotes}
\end{threeparttable}
\end{table}

\section{Screenshots of HelixVS Server}
\label{sec:screenshots}

\begin{figure}
\centering
\includegraphics[width=0.9\columnwidth]{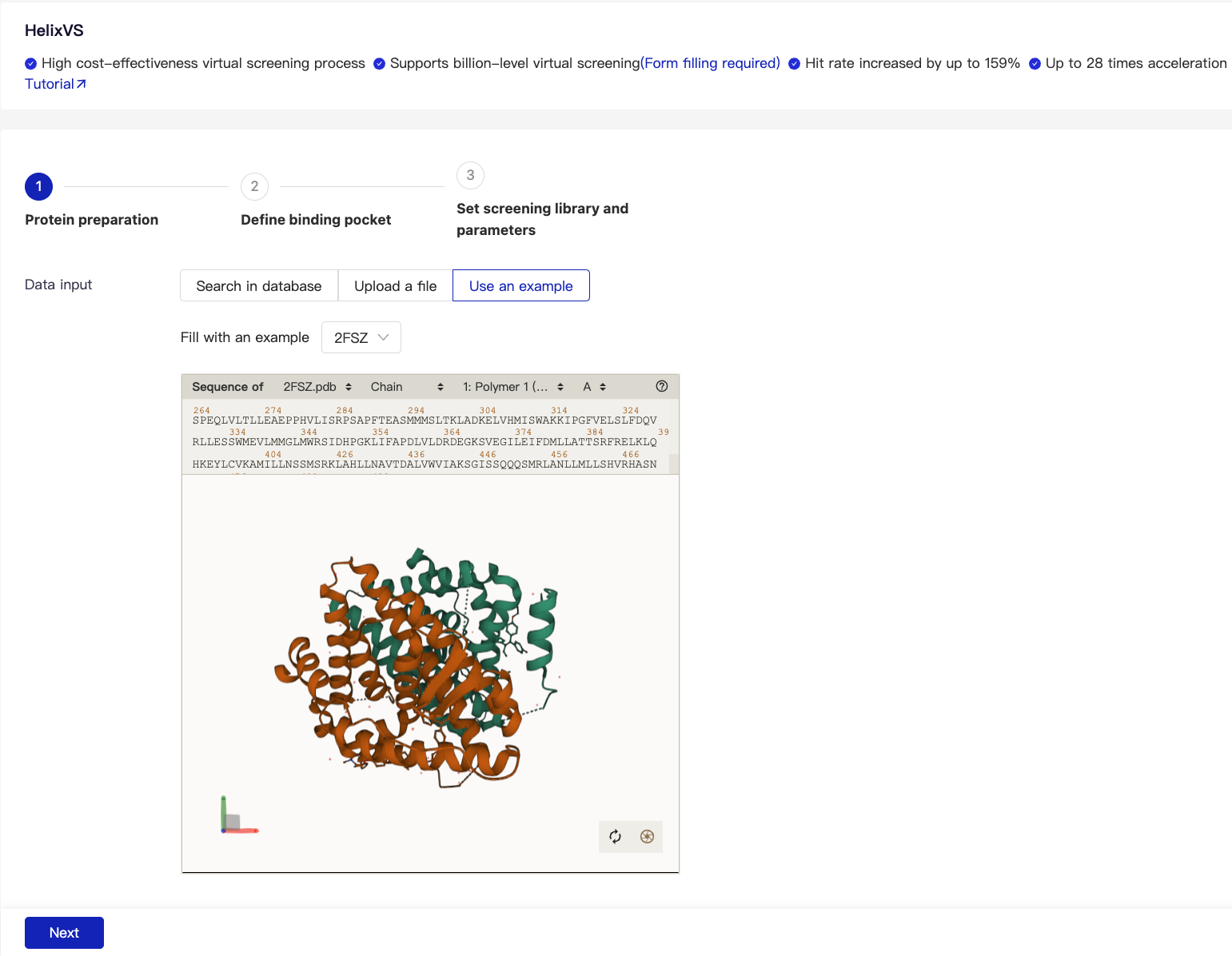}
\caption{The protein preparation page. The protein structure can be uploaded in PDB format or downloaded from PDB database. This is the same for both HelixVS and HelixVS-Syn.}
\label{fig:helixvs_server_prot}
\end{figure}

\begin{figure}
\centering
\includegraphics[width=0.9\columnwidth]{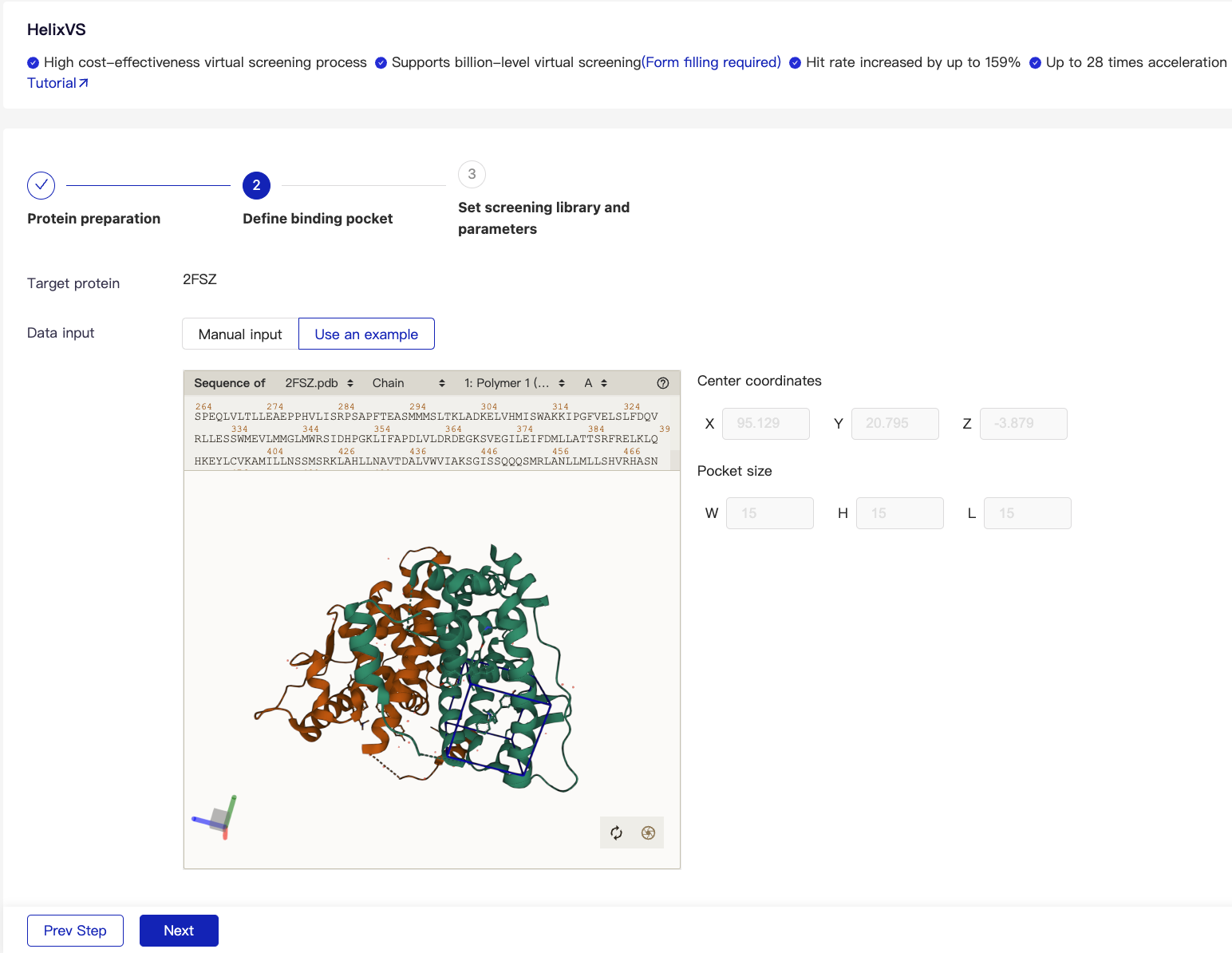}
\caption{The pocket specification page. The pocket can be specified by setting a box with center coordinates and size on three dimensions in Angstrom. The pocket will be displayed on the protein structure by clicking the "Show Pocket" button. This is the same for both HelixVS and HelixVS-Syn.}
\label{fig:helixvs_server_pocket}
\end{figure}

\begin{figure}
\centering
\includegraphics[width=0.9\columnwidth]{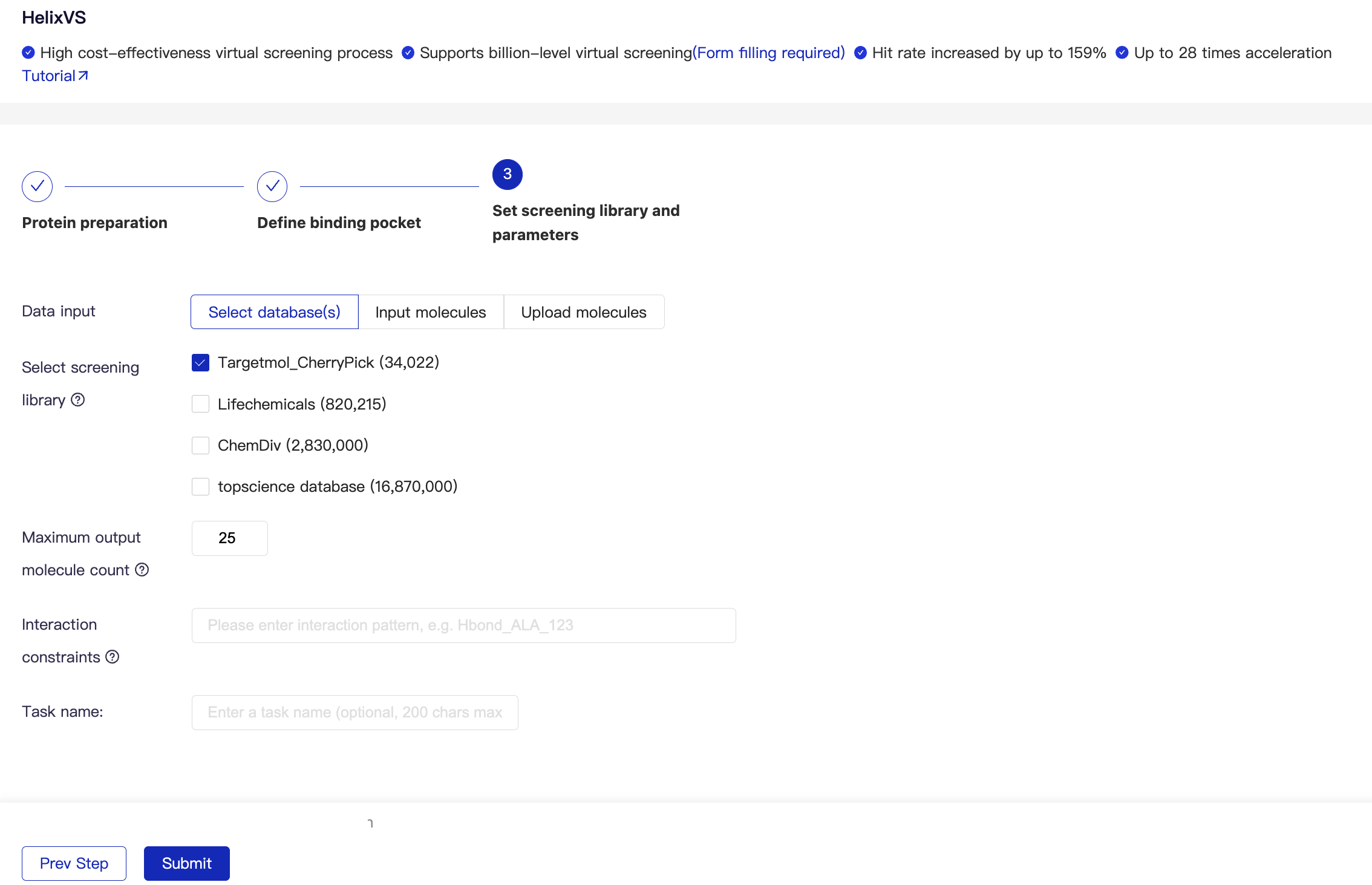}
\caption{The parameter setting page of HelixVS. The user can set the built-in screening library, and the interaction constraints.}
\label{fig:helixvs_server_param}
\end{figure}

\begin{figure}
\centering
\includegraphics[width=0.9\columnwidth]{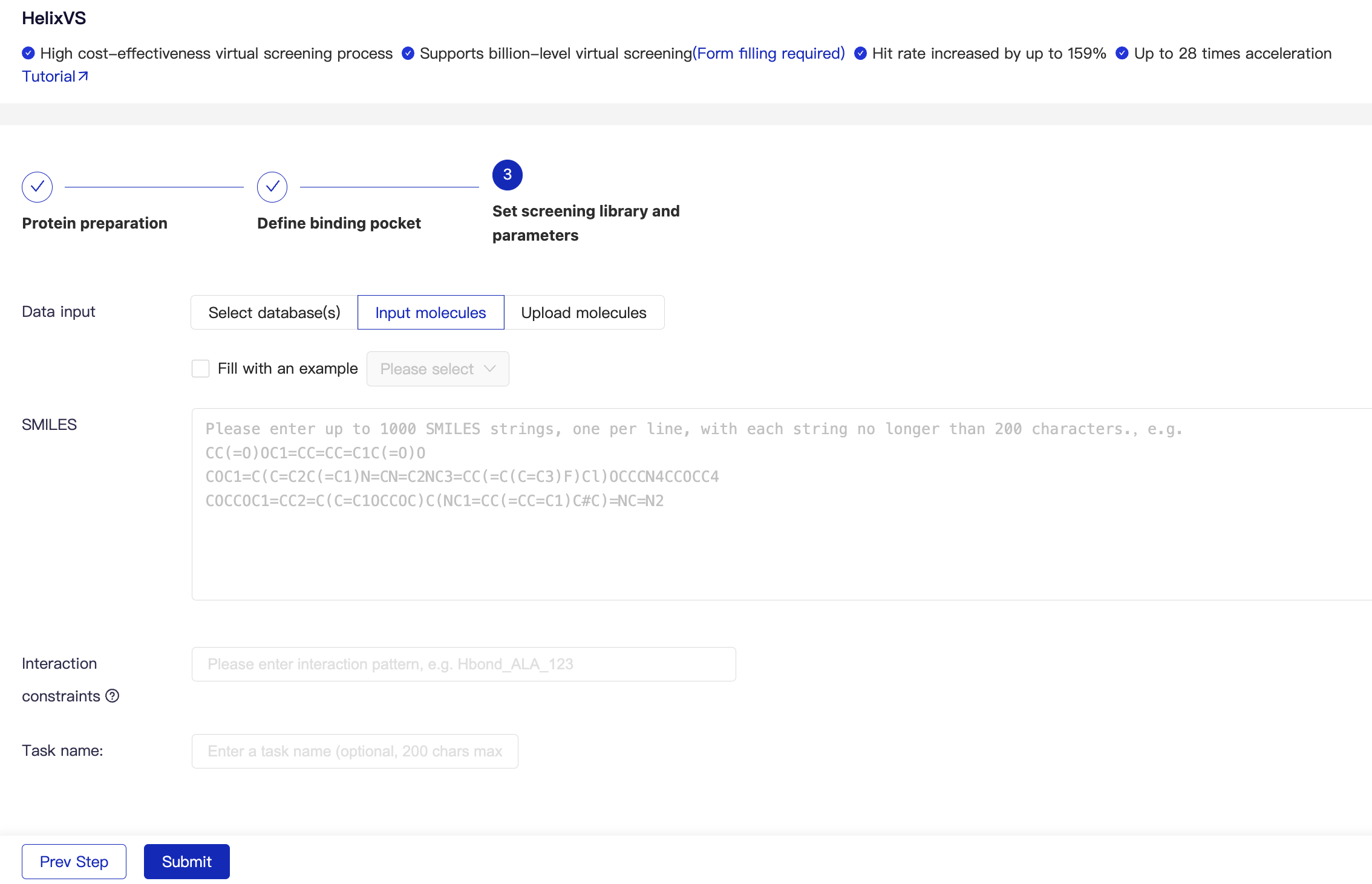}
\caption{The parameter setting page of HelixVS. The user can input molecules as the screening library and set the interaction constraints.}
\label{fig:helixvs_server_param_2}
\end{figure}

\begin{figure}
\centering
\includegraphics[width=0.9\columnwidth]{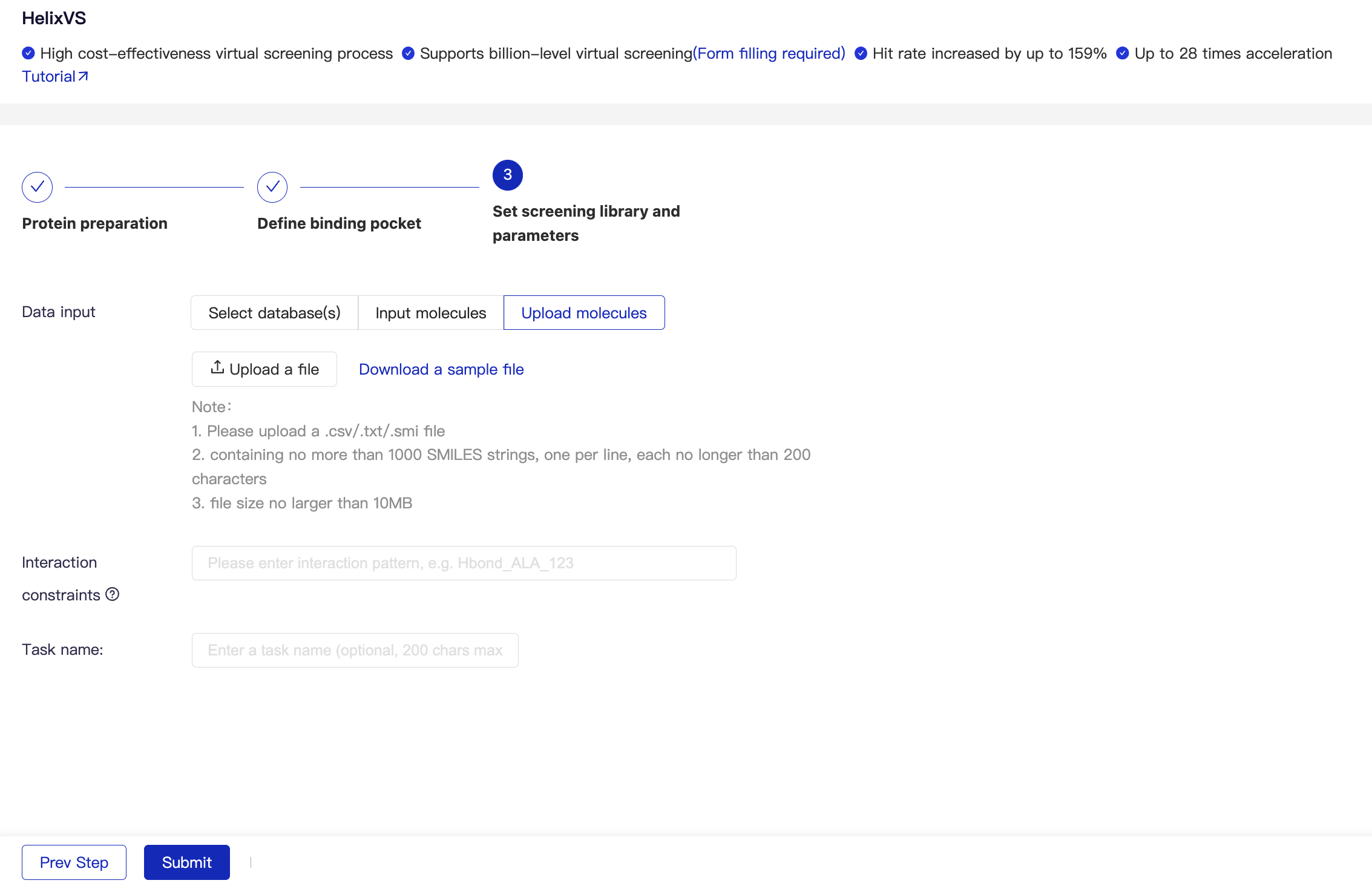}
\caption{The parameter setting page of HelixVS. The user can upload a molecule file as the screening library and set the interaction constraints.}
\label{fig:helixvs_server_param_3}
\end{figure}

\begin{figure}
\centering
\includegraphics[width=0.9\columnwidth]{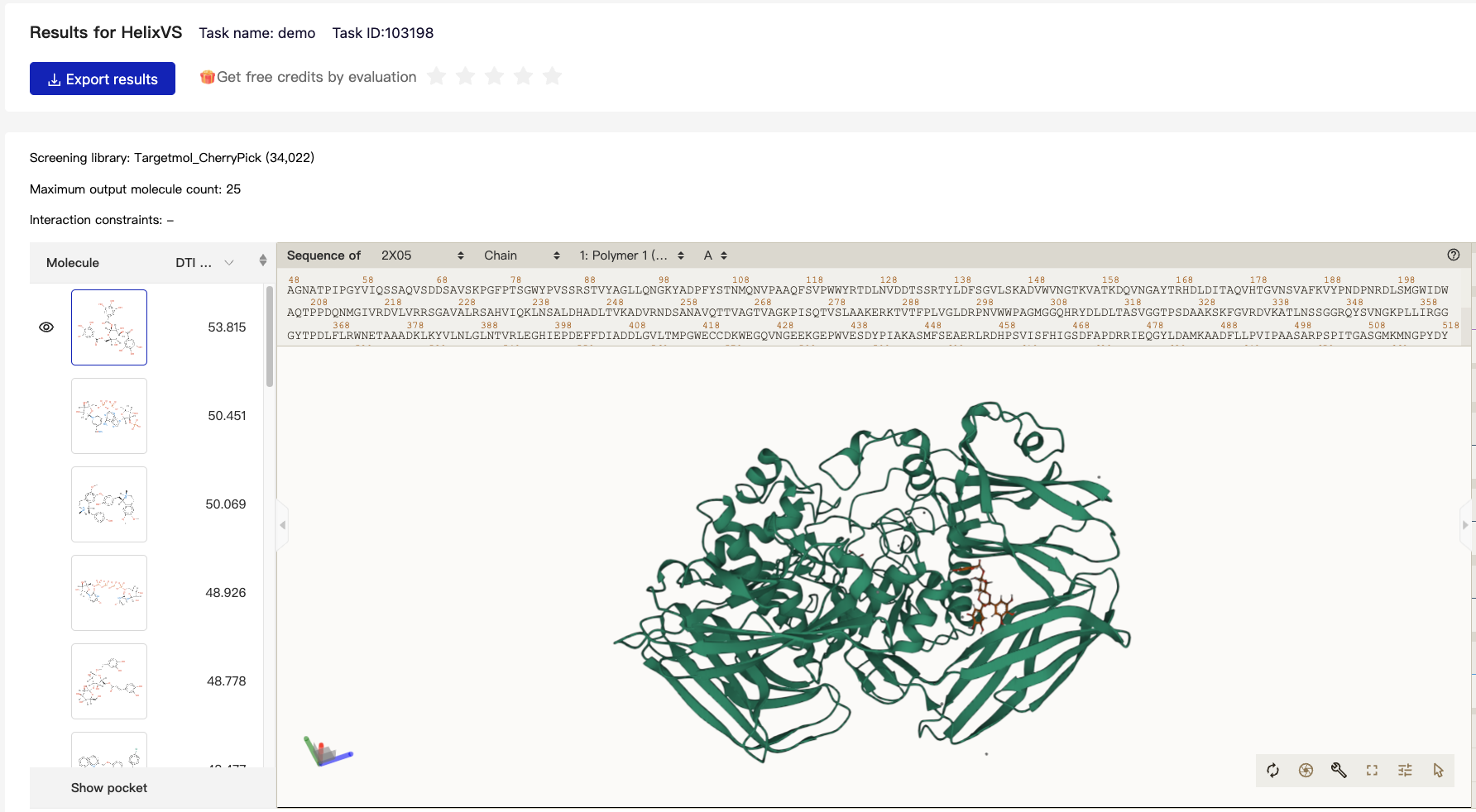}
\caption{The result page showing the screening task results. The left panel displays the filtered molecular list and 2D topological molecular structures, while the right panel shows the binding structure of the currently selected molecule with the target protein. Structure visualization is implemented using Mol* Viewer \cite{sehnal_mol_2021}.}
\label{fig:helixvs_server_result}
\end{figure}

\begin{figure}
\centering
\includegraphics[width=0.9\columnwidth]{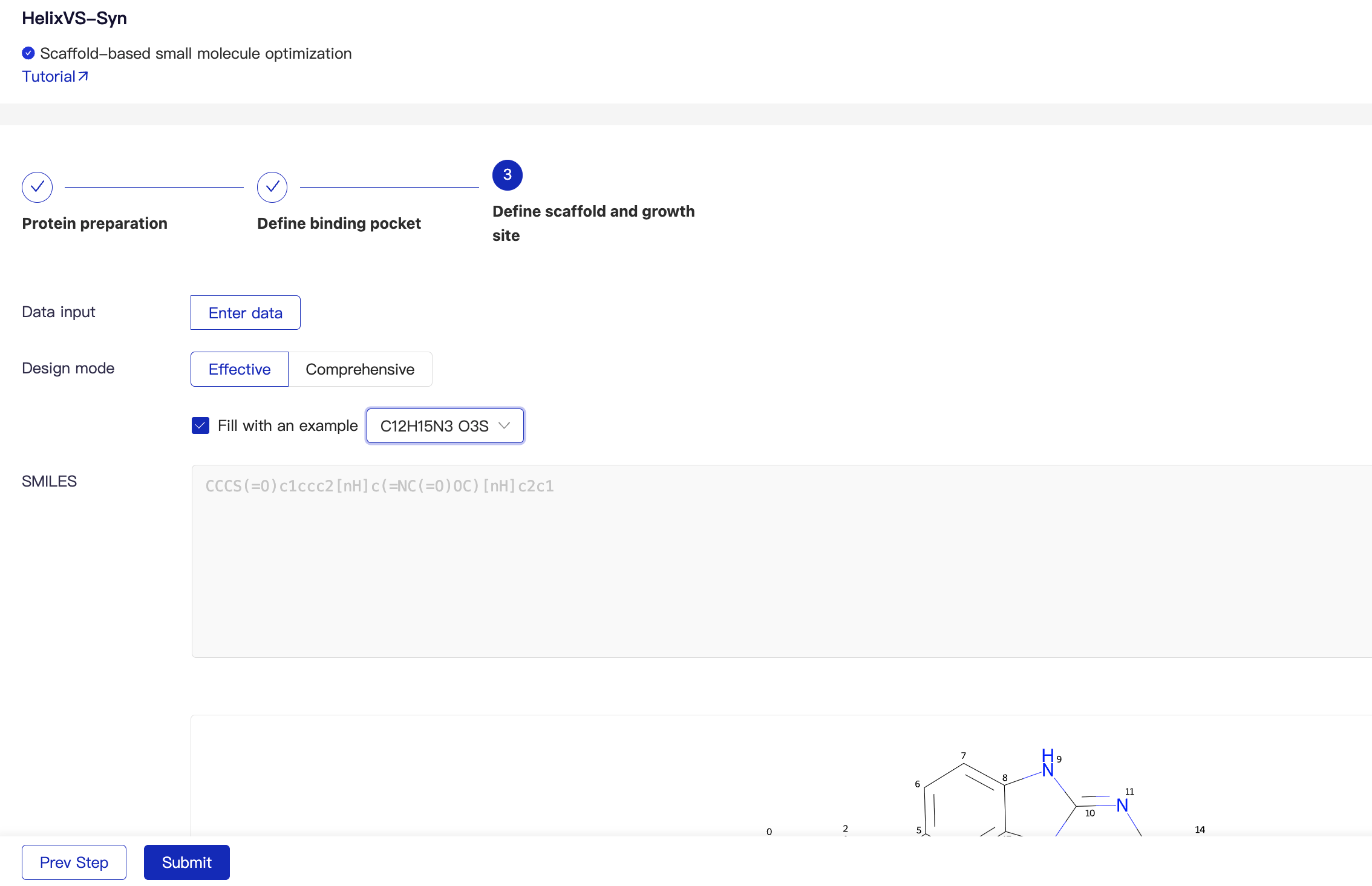}
\caption{The parameter setting page of HelixVS-Syn(1). The user can set the design mode and reference molecule.}
\label{fig:helixvs_server_param_syn_1}
\end{figure}

\begin{figure}
\centering
\includegraphics[width=0.9\columnwidth]{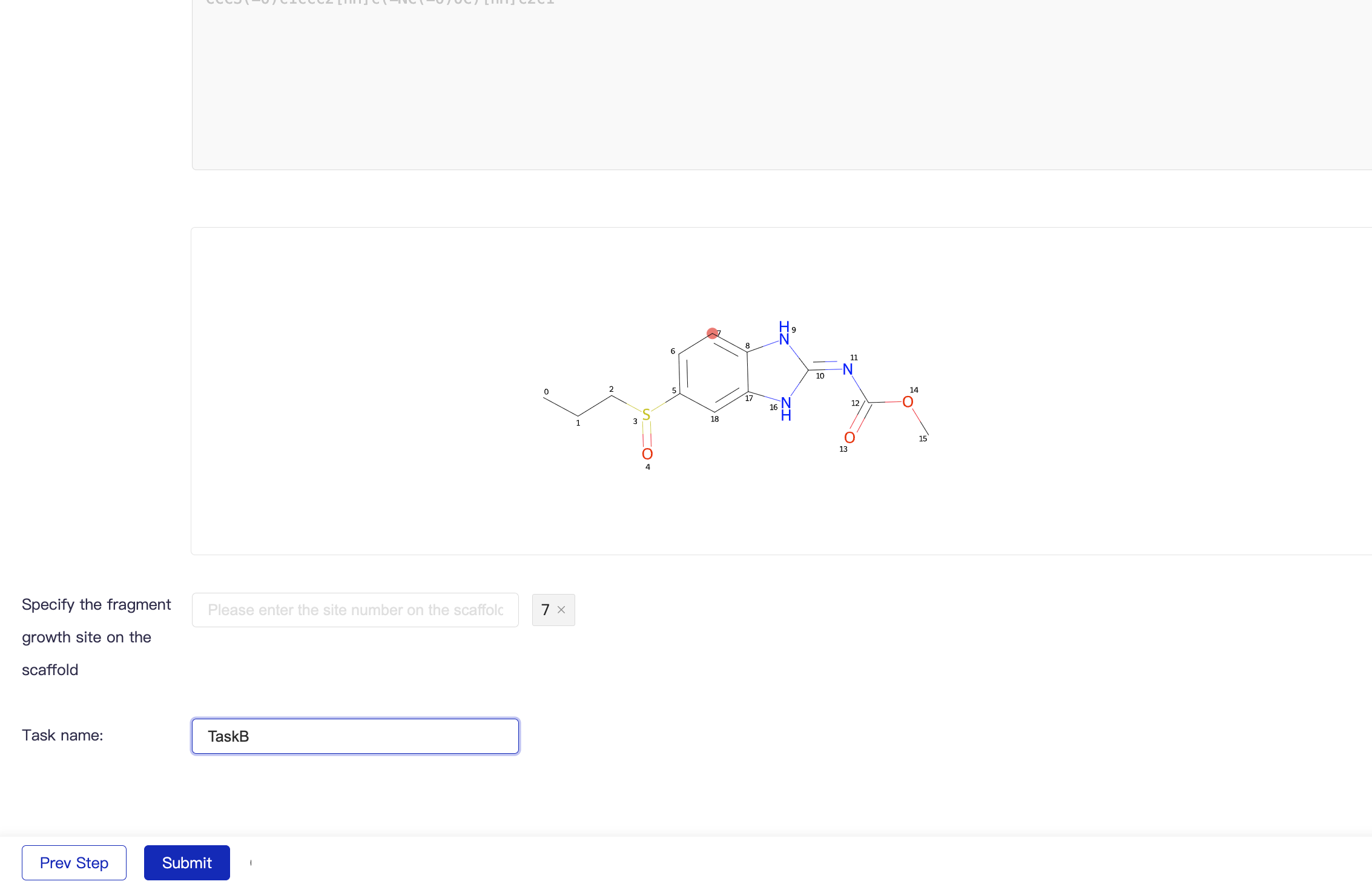}
\caption{The parameter setting page of HelixVS-Syn(2). The user can set the fragment growth site on the scaffold and the interaction constraints.}
\label{fig:helixvs_server_param_syn_2}
\end{figure}

\begin{figure}
\centering
\includegraphics[width=0.9\columnwidth]{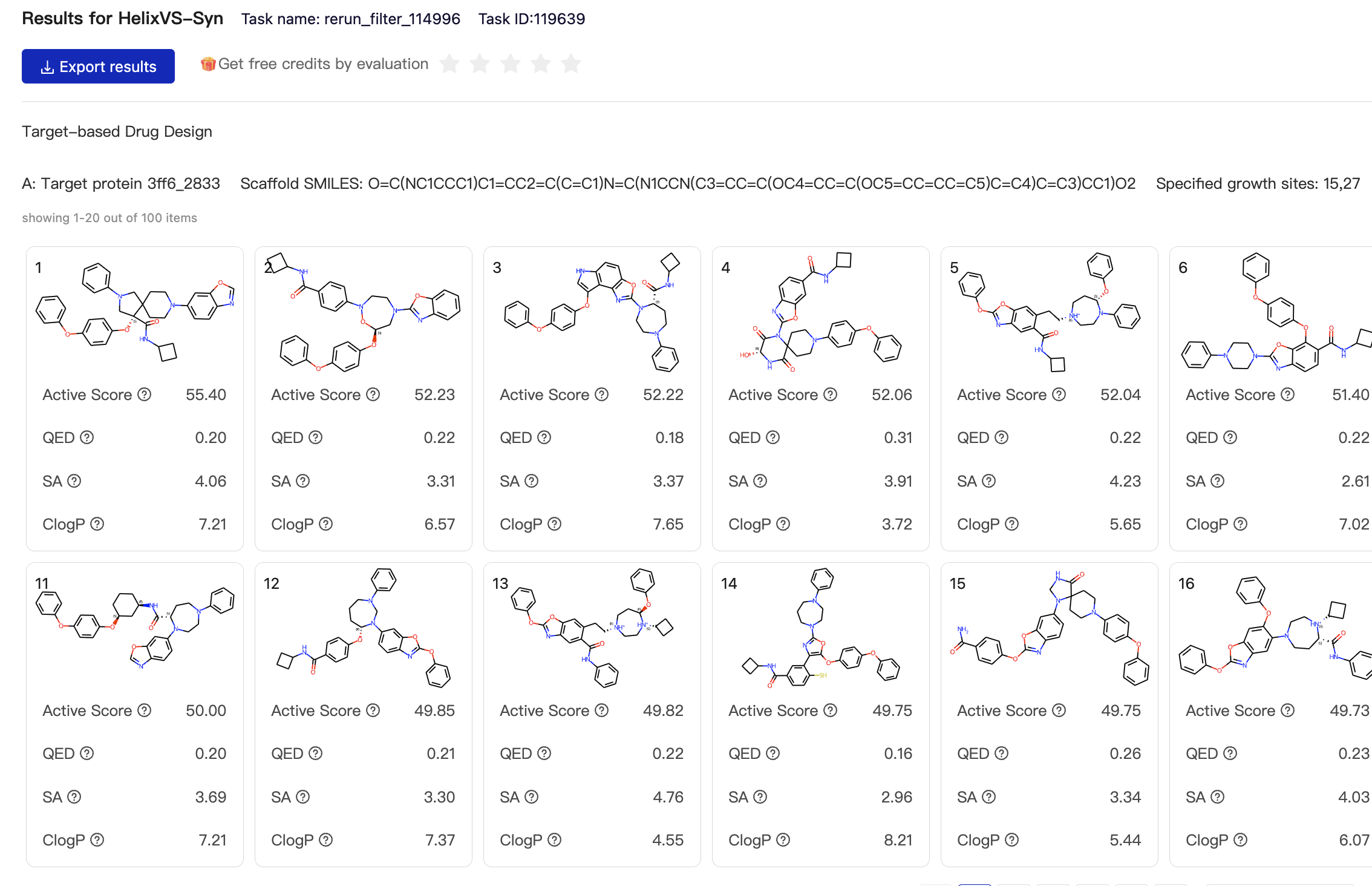}
\caption{The card view of the result page of HelixVS-Syn. The user can see the 4 properties of the output molecules: Active Score, QED, SA and ClogP. }
\label{fig:helixvs_server_result_syn_1}
\end{figure}

\begin{figure}
\centering
\includegraphics[width=0.9\columnwidth]{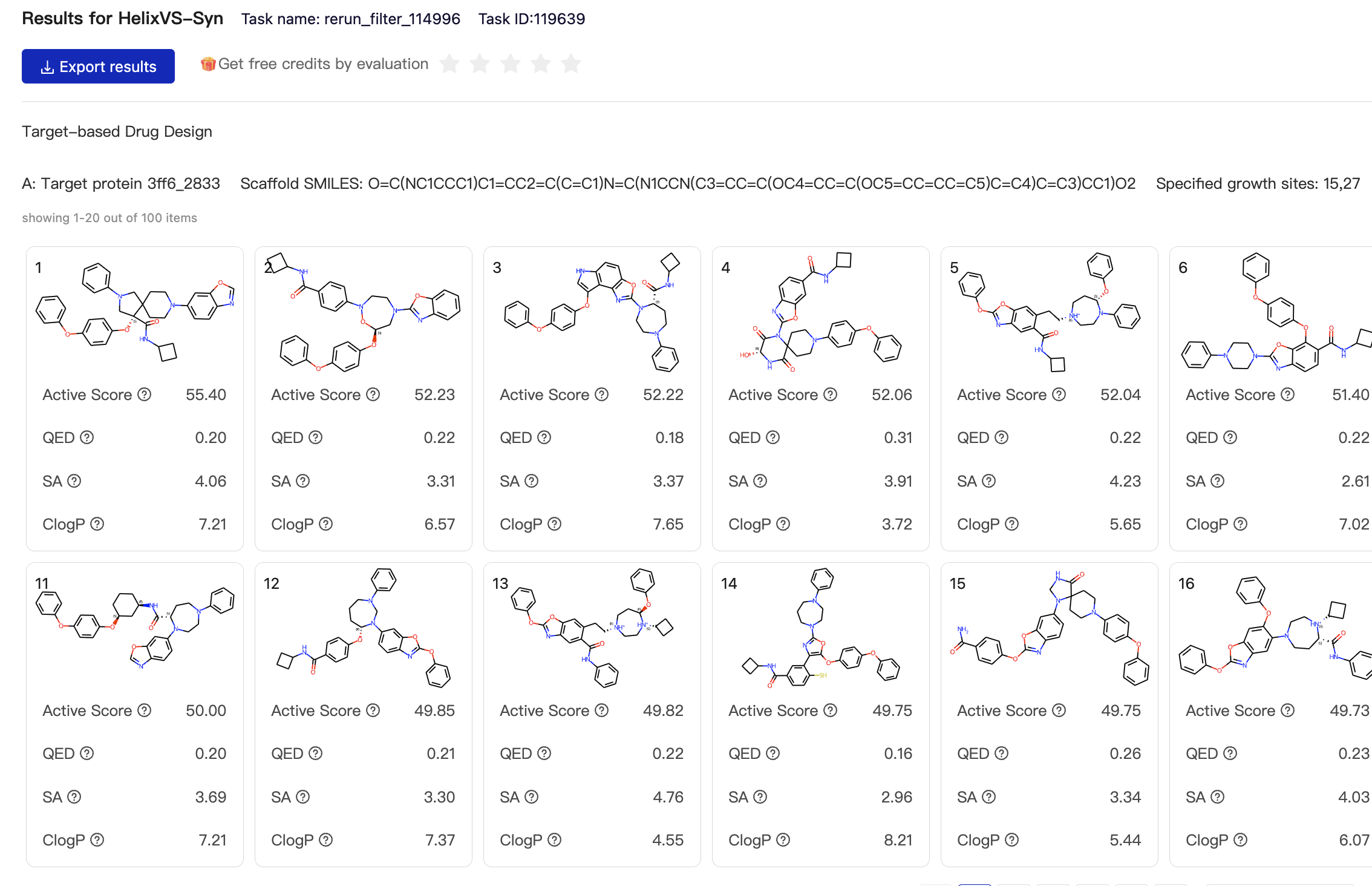}
\caption{The 3D view of the result page of HelixVS-Syn. This page is the same as the HelixVS result rage.}
\label{fig:helixvs_server_result_syn_2}
\end{figure}






\end{appendices}
\clearpage

\bibliographystyle{unsrtnat}
\bibliography{references}  

\end{document}